\documentclass[prb,twocolumn,superscriptaddress,showpacs,amsmath,amssymb]{revtex4}

\usepackage{graphicx}
\usepackage{latexsym}
\usepackage{amsmath}
\usepackage{amssymb}
\usepackage{amsfonts}
\usepackage{color}
\usepackage{bm}
\usepackage{verbatim}
\usepackage{color}

\begin{document}

\title{Stable fractional flux vortices and unconventional magnetic state in two-component superconductors.}

\author{Mihail Silaev}

 \affiliation{Institute for Physics of Microstructures RAS, 603950
Nizhny Novgorod, Russia.}
\affiliation{Department of Theoretical Physics, The Royal
Institute of Technology, Stockholm, SE-10691 Sweden}

\begin{abstract}
In the framework of London theory we study the novel magnetic
state in two-component superconductors with finite density of
fractional flux vortices stabilized near the surface. We show that
the process of vortex entry into the two-component superconductor
consists of several steps, while the external magnetic field
increases from zero. At the first stage only vortices in one of
the order parameter components penetrate and sit at the
equilibrium position near the surface. When the magnetic field is
increased further vortices in the second order parameter component
eventually enter the superconductor. Such a complex partial vortex
penetration leads to the modification of a Bean-Livingston barrier
and a magnetization curve as compared to conventional
single-component superconductors. We discuss the possibility of
experimental identification of protonic superconductivity in the
projected superconducting state of liquid metallic hydrogen and
hydrogen rich alloys with the help of partial vortex penetration
effect.
\end{abstract}

 \maketitle

\section{Introduction}

Recently there has been a considerable interest to the physics of
superconductors with multiple order parameter components, renewed
by the discovery of a number of two-band superconducting
materials, such as $MgB_2$ \cite{MgB2},
ferropnictides\cite{Ferropnictides}, heavy fermion compounds
\cite{HeavyFermion}, borocarbides\cite{Borocarbides} and the
 advances in experimental techniques\cite{Experiment} providing an intriguing
 possibility of observing new macroscopic quantum phenomena in
 liquid metallic hydrogen at ultra high pressures\cite{Hydrogen1,Hydrogen2, Hydrogen3}.
Historically the first realization of two component
superconducting state was considered by
Moskalenko\cite{Moskalenko} and by Suhl, Matthias, and
Walker\cite{Suhl} in a metal with two overlapping energy bands on
the Fermi surface.

Multiple-component superconductors feature qualitatively new
effects, with respect to the conventional
ones\cite{Babaev-VorticesNature}. One of the striking differences
is an unconventional structure of the mixed state in such
superconductors. For example, in
Ref.\onlinecite{Babaev-SemiMeissner} it was demonstrated that in
two-band superconductor the interaction between vortices of the
equal vorticity is not necessarily purely repulsive, but in some
cases may be characterized by long-range attraction and
short-range repulsion. This in particular leads to the
clusterization of vortices and formation of the "semi-Meissner
state", which was recently claimed to be experimentally observed
in $MgB_2$ \cite{Moshchalkov1-5}.

Also a rich variety of peculiar vortex states in two-component
superconductors have been investigated recently in the framework
of the extended Ginzburg-Landau (GL) theory\cite{Babaev-GL}. In
particular, there was found a class of solutions of GL equations
describing vortices bearing a fractional number of magnetic flux
quanta. In general, such vortices can be characterized by
different winding numbers of the order parameter in the two
superconducting components, i.e. $L_1\neq L_2$. Among such
vortices one can distinguish a subclass of vortices which have
$L_1\neq0$ and $L_2=0$, when the phase winding exists only in one
of the condensates. In the present paper we will focus on the
simplest case when $L_1=\pm 1$ and $L_2=0$ (or equivalently
$L_2=\pm 1$ and $L_1=0$), further using the term "fractional
vortices" to identify such objects. Being positioned at the same
point the two fractional vortices in different condensates
characterized by the equal vorticity form a composite vortex
bearing a single quantum of magnetic flux.

The fractional vortices in two-component superconductors are
qualitatively different from thermodynamically stable exotic
topological defects that occur in superfluid and superconducting
systems with multiple order parameter components, such us
superfluid $^3$He (see the review Ref. \onlinecite{VolovikRMP}),
ultracold atoms \cite{ColdAtoms} or vortices in d-wave HTSC
cuprates with induced s-wave order parameter component inside
vortex core\cite{HTSC}. The fractional vortices considered in the
present work are thermodynamically unstable in bulk two-component
superconductor\cite{Babaev-GL} since their energy per unit length
is logarithmically or linearly divergent with the sample size.
Therefore it is impossible to create the fractional vortices in
bulk superconductor by applying the external magnetic field.
 This is the reason why up to date
fractional vortices in two component superconductors have not been
observed. However it was proposed that the deconfinement of
fractional vortices forming a composite vortex can occur due to
the thermal fluctuations\cite{ThFluct1, ThFluct2} or due to the
thermal creation of fractional vortex-antivortex
pairs\cite{Babaev-NuclPhys}. Also, with the help of GL
calculations fractional vortices were demonstrated to exist in
mesoscopic two-component superconductors\cite{Chibotaru, Peeters}.
In the present paper we propose that fractional vortices can be
thermodynamically stable near the surface of superconductor and
therefore there is a possibility to create a finite density of
them by external magnetic field. The finite density of fractional
vortices near the surface forms a novel magnetic state of
superconductor which should be manifested by a modification of
magnetization curve.

Further we provide an analytical treatment of London equations
describing a behaviour of vortices near the surface of a
two-component superconductor. We consider the two component
superconductor as a mixture of two individually conserved
superconducting condensates. This model is relevant for the
systems where the interband coupling is forbidden by symmetry.
Among such systems currently generating great interest is the
liquid metallic hydrogen at ultra high pressures\cite{Experiment}.
In this case two superconducting components were predicted to
originate from electronic and protonic Cooper pairing in metallic
hydrogen and hydrogen rich alloys \cite{Hydrogen1, Hydrogen2,
Hydrogen3}. Since the Cooper pairs of electrons cannot be
converted into the Cooper pairs of protons the interband Josephson
interaction is strictly zero. An analogous model of two
individually conserved condensates was considered recently in
order to describe the exotic states of matter at neutron star
inner cores with several charged barionic components, namely
$\Sigma^{-}$ hyperons and protons. The mixture of superfluid
$\Sigma^{-}$ hyperons and protons is analogous to the two-gap
superconductor with strictly zero interband Josephson coupling
\cite{NeutronStar1, NeutronStar2}.

On the other hand in two-band superconductors where the Cooper
pairing of electrons takes place in different bands
\cite{Moskalenko, Suhl} the superconducting condensates in general
cannot be considered as individually conserved. However the model
with negligible interband Josephson coupling can also be applied
to describe multiband superconductors provided that all relevant
physical scales are much shorter than the Josephson length.
Although basically considering the system without the interband
Josephson coupling we will discuss on a qualitative level the
modification of our results due to this effect.

It is well known that the entry of vortices into a type-II
superconductor is hindered by the so-called Bean- Livingston
surface barrier \cite{BeanLivingston, Schmidt}. This barrier
arises due to a competition of two forces acting on the vortex
line: a force coming from the Meissner current driving the vortex
into the superconductor and a force of the vortex' mirror image
attracting it towards the outside. As a result, the penetration
field $H_s$ of first vortex entry is typically much larger than
the lower critical field $H_{c1}$ and becomes of the order of the
thermodynamic critical field $H_c$ \cite{Schmidt}. Recently it was
demonstrated that a distribution of the Meissner current and the
value of the penetration field $H_s$ are highly sensitive to the
presence of Andreev bound state at the surface of $d$-wave
superconductors\cite{Schopohl}. It is the goal of the present
paper to show that the two-component structure of the order
parameter can also alter significantly the value of the
penetration field $H_s$ and the process of the vortex entry in
two-gap superconductors.

In two-component superconducting systems it is natural to expect
that the penetration of fractional vortices in the component with
lower condensation energy should occur at magnetic field which is
smaller than $H_c$. However, since the existence of fractional
vortices is prohibited in the bulk, they should sit at a certain
distance from the surface, corresponding to the energy minimum.
The further increase of magnetic field will lead to the nucleation
of vortices in the second superconducting component. At a final
stage of the process of vortex entry the fractional vortices of
different types merge and proliferate into the bulk.

This paper is organized as follows. In Sec.~\ref{sec:theory} we
give an overview of the theoretical framework, namely the London
theory of two-component superconductors. In Sec.~\ref{sec:results}
we discuss the Gibbs energy of fractional
 vortices near the surface and address the questions of fractional
 vortex stability and the Bean-Livingston barrier for penetration of fractional
 vortices. Also we calculate the equilibrium distribution of
 fractional vortices near the surface.
 In Sec.~\ref{sec:discussion} we discuss the modification of the
 magnetization curve due to the partial vortex
 penetration and the possibility to implement the experimental identification
 of two order parameter components in the projected superconducting state of
 liquid metallic hydrogen. We give our conclusions in
Sec.~\ref{sec:summary}.


\section{Basic equations: London theory of fractional vortices in two-gap superconductor}
\label{sec:theory}


 Let us consider a superconductor with two coexisting  superconducting condensates.
   The external magnetic field is directed along the $z$ axis ${\bf H_0}=H_0 {\bf z_0}$.
    Considering the London limit, i.e.
assuming that the coherence length is vanishingly small compared
to all other length scales we obtain the Gibbs energy per unit
length along the $z$ axis as follows \cite{ThFluct2}:
 \begin{equation}\label{Gibbs0}
 F_G=F-\frac{1}{4\pi}\int {\bf H}\cdot{\bf H_0} d^2r,
 \end{equation}
 where the free energy is
\begin{eqnarray}\label{FreeEnergy}
  F&=&\frac{1}{8\pi}\int  \left[ {\bf H}^2+\lambda_A^{-2}\left({\bf A}-
  \frac{\phi_0}{2\pi}\nabla \theta_A\right)^2\right.\\
  \nonumber
  &+&\lambda_B^{-2}\left.
   \left({\bf A}-\frac{\phi_0}{2\pi}\nabla
 \theta_B\right)^2\right] d^2r.
\end{eqnarray}
 Here $\lambda_{A,B}$ are the different length scales,
 proportional to the densities of two types of superconducting
 carriers, ${\bf H}$ is a total magnetic field, ${\bf A}$ is a
 vector potential, and $\theta_{A,B}$ are the phases of
 superconducting order parameters. As we will see further, the London limit in
 two-component model that we consider yields an exponential decaying magnetic
 field generated by fractional vortices.
 On the other hand it was recently demonstrated that in full two-component GL
 theory there is always a power-law tail of magnetic field around
 a fractional vortex \cite{Babaev-delocalization}. However the power law behaviour of
 magnetic field starts at the distances greater than the
 London penetration depth $\lambda$ which is considered to be the
 largest length scale in our present paper. For example, as we will see below
 the intervortex distance in the proposed novel magnetic state of two-component
 superconductor is much smaller than $\lambda$. Within this parameter range
 the London model is a good approximation to study the magnetic properties of two-component
 superconductors.

 The London- Maxwell (LM) equation for the magnetic field
 obtained from the condition $\delta F/\delta {\bf A}=0$ yields:
\begin{equation}\label{London-Maxwell}
  \lambda^2 \nabla\times {\bf H}=-{\bf A}
 +\frac{\phi_{A}}{2\pi}\nabla \theta_A+\frac{\phi_{B}}{2\pi}\nabla
 \theta_B,
\end{equation}
 where $ \lambda^2=\lambda_A^2\lambda_B^2/(\lambda_A^2+\lambda_B^2)$
 is the London penetration depth, and
 $ \phi_{A,B}=\phi_0(\lambda/\lambda_{A,B})^2$ are partial magnetic fluxes so that $\phi_{A}+\phi_{B}=\phi_0$.
 With the help of LM equation (\ref{London-Maxwell}) the free energy (\ref{Gibbs0}) can
 be rewritten as a superposition of two parts
 \begin{equation}\label{Division}
  F=F_m+F_{nc},
 \end{equation}
 where the first part is the sum of total magnetic energy and the
 kinetic energy of superconducting current and thus depends only
 on the magnetic field
 \begin{equation}\label{Fm}
 F_m=\frac{1}{8\pi}\int\left[{\bf H}^2+ \lambda^2 (\nabla\times
 {\bf H})^2\right]  d^2r.
 \end{equation}
 The second part $F_{nc}$ is the kinetic energy of relative motion
 of two condensates, which can be associated with the neutral
 superfluid current, supported by the counterdirected motion of
 equally charged particles (electrons in two-gap superconductors) or
 codirected motion of oppositely charged ones (electrons and protons in superconducting liquid metallic
 hydrogen) \cite{Babaev-VorticesNature}. This term depends on the relative phase difference
 between two condensates $\varphi_{rel}=\theta_A-\theta_B$ as
 follows
 \begin{equation}\label{Frel}
  F_{nc}=\frac{1}{2\pi}\frac{\phi_A\phi_B}{(4\pi\lambda)^2}\int \left(\nabla \varphi_{rel}\right)^2
  d^2r.
 \end{equation}
  Such division of the total energy (\ref{Division})
 is convenient for the calculations since the parts
 $F_m$ and  $F_{nc}$ can be
 found independently.

 For example the energy $F_m$ of a one fractional vortex in $A$ or $B$ condensate
 can be calculated exactly in a same scheme as for the case of conventional
 single gap superconductor \cite{Schmidt}.
 The result is
  \begin{equation}\label{Fm1-0}
  F_m=\varepsilon_i+\frac{\phi_{i}}{8\pi} H_{v}({\bf R_{i}}),
 \end{equation}
 where $i=A,B$ and ${\bf R_{i}}$ is the fractional vortex coordinate. The first term in Eq.(\ref{Fm1-0}) is the
 is the self energy of the vortex
 $$
 \varepsilon_i=\left(\frac{\phi_{i}}{4\pi \lambda}\right)^2 \ln
 (\lambda/\xi_{i}),
 $$
 where $\xi_i$ is the coherence length and the second term in Eq.(\ref{Fm1-0}) is the energy in the
  magnetic field ${\bf H_{v}}=H_{v}{\bf z_0}$ generated by
  other vortices.

 At the infinite superconductng sample the part of the energy
  $F_{nc}$ given by (\ref{Frel}) is divergent
 for a single fractional vortex due to the unscreened currents induced in the vortex free phase \cite{Babaev-GL}.
 Indeed, taking for example $\theta_A=\arctan(y/x)$ and $\theta_B=0$
 we get that $(\nabla \varphi_{rel})^2\sim 1/r^2$ which means that
 the expression (\ref{Frel}) is logarithmically divergent with
  the size of the superconducting sample $L$ so that $F_{nc}\sim \ln
  L$. However further we will see that in some specific cases the
 energy of a single fractional vortex can be finite.
  In  particular such situation is realized for fractional vortex placed near the boundary of superconductor.
  In this case the large scale divergence in Eq.(\ref{Frel}) is removed due to the cancellation of the
  unscreened superconducting current in vortex free phase due to the image antivortex.


\section{Results}\label{sec:results}

\subsection{Gibbs energy of fractional vortices near the surface of superconductor.}
\label{subsec:GibbsEnergy}

\begin{figure}[htb]
\centerline{\includegraphics[width=1.0\linewidth]{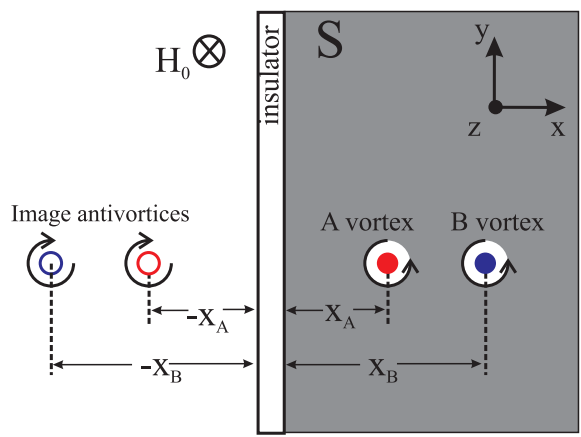}}
\caption{\label{Fig:model} The sketch of the system under
consideration. It consists of a two - component superconductor,
occupying the half space $x>0$ bounded by the $yz$ plane. The
fractional vortices are situated at the distances $x_A$ and $x_B$
from the boundary. The boundary conditions are taken into account
by placing image antivortices at the proper points behind the
boundary plane.
  }
\end{figure}

The sketch of the system considered is shown in Fig.
\ref{Fig:model}. It consists of a two- component superconductor,
occupying the half space $x>0$ bounded by the $yz$ plane. Our goal
is to calculate the Gibbs energy (\ref{Gibbs0}) of the vortex
configuration shown in Fig.1, i.e. of the two vortices in $A$ and
$B$ phases located at the points ${\bf R_A}=(x_A,0 )$ and ${\bf
R_B}=(x_B,0)$ near the surface. Here we can consider only vortices
positioned in a line since any relative shift of $A$ and $B$
vortices along the boundary plane leads to the energy rising.

 All vortices are directed along the $z$ axis, which is
 parallel to the boundary plane. Since not Meissner currents nor
 vortices parallel to the surface do not create magnetic fields
 outside the superconducotr\cite{Schmidt},
 the integration in Eq.(\ref{Gibbs0}) can be restricted
 to the superconducting region only. Furthermore, in Eq.(\ref{FreeEnergy}) we can
 take into account only the magnetic field generated by vortices
 and neglect the field generated by the Meissner current.
 Then to calculate the energy we should find the magnetic field
 ${\bf H} ({\bf r})$ generated by vortices and the relative phase
 distribution $\varphi_{rel}({\bf r})$.

The magnetic field is determined by the Maxwell-London equation
having the following form
\begin{align}\label{London-Maxwell1}
&\lambda^2\nabla\times\nabla\times {\bf H}+{\bf H}=\\ \nonumber
&{\bf z_0}\left[\phi_{A} \delta ({\bf r}-{\bf
 R_{A}})+\phi_{B}\delta ({\bf r}-{\bf
 R_{B}})\right].
\end{align}
 Varying the expression (\ref{Frel}) with respect to
the phase difference $\varphi_{rel}$ we obtain the following 2D
Poisson equation with the sources at the points of vortex
positions
\begin{equation}\label{sine-Gordon}
  \Delta\varphi_{rel}= 2\pi\left[\delta ({\bf r}-{\bf R_{A}})-\delta ({\bf r}-{\bf
   R_{B}})\right].
 \end{equation}

The Eqs.(\ref{London-Maxwell1},\ref{sine-Gordon}) should be
supplemented by the boundary conditions at the surface of
superconductor
 \begin{equation}\label{BC}
 {\bf x_0}\cdot\left({\bf A}-\frac{\phi_0}{2\pi}\nabla
 \theta_{A,B}\right)|_{x=0}=0,
 \end{equation}
 which means vanishing of the superconducting current through the
 boundary in both of the condensates.
These boundary problems can be treated by using the method of
images, i.e. by placing the image anti-vortices in $A$ and $B$
phases at the points ${{\bf \tilde{R}_{A,B}}}=(-x_{A,B},0)$
respectively [see Fig.(\ref{Fig:model})]. Then, the free energy of
the two-vortex molecule near the flat surface of superconductor is
just a one half of the free energy of vortex-antivortex molecule
[Fig.(\ref{Fig:model})].

Let us evaluate the energy $F_m$ given by the
 expression (\ref{Fm}). For the stack of several vortices it is given by the
 sum of the individual vortex energies (\ref{Fm1-0}).
  The magnetic field ${\bf H}=H_v {\bf z_0}$ generated by the
 vortex currents can be taken by the superposition of fields
 produced by vortices at the points ${\bf R_{A,B}}$ and
 antivortices at points ${\bf \tilde{R}_{A,B}}$:
 $$
 H_v=H_{v1}+H_{v2}+ H_{av1}+H_{av2},
 $$
 where
 $$
 H_{v1}= \frac{\phi_{A}}{2\pi \lambda^2}
 K_0\left(\frac{|{\bf r - R_{A}}|}{\lambda}\right),
 $$
 $$
 H_{v2} =\frac{\phi_{B}}{2\pi \lambda^2}
 K_0\left(\frac{|{\bf r - R_{B}}|}{\lambda}\right),
 $$
 $$
 H_{av1} =-\frac{\phi_{A}}{2\pi \lambda^2}
 K_0\left(\frac{|{\bf r - \tilde{R}_{A}}|}{\lambda}\right),
 $$
 $$
 H_{av2} =-\frac{\phi_{B}}{2\pi \lambda^2}
 K_0\left(\frac{|{\bf r - \tilde{R}_{B}}|}{\lambda}\right),
 $$
 where $K_0(x)$ is the zero - order  Hankel function, having the
asymptotic $K_0(x)\approx C-\ln(x)$ at $x\ll 1$ and $K_0(x)\approx
\sqrt{\pi/2x}e^{-x}$ at $x\gg 1$ (Ref.\onlinecite{Abramoviz}).
 Thus for the energy $F_m$ we obtain:
 \begin{align}\label{Fm2}
  &F_m=\varepsilon_A+\varepsilon_B+
 \frac{\phi_{A} \phi_{B}}{(4\pi\lambda)^2 }
 \left[ 2K_0\left(\frac{|x_A-x_B|}{\lambda}\right)-\right. \\
 \nonumber
 &\left. 2K_0\left(\frac{|x_A+x_B|}{\lambda}\right)-
 \sigma K_0\left(\frac{2x_A}{\lambda}\right)
 -\frac{1}{\sigma}K_0\left(\frac{2x_B}{\lambda}\right)\right],
 \end{align}
 where we introduce the coefficient $\sigma=\phi_A/\phi_B$.

To calculate the other part of free energy $F_{nc}$ we should
substitute into Eq.(\ref{Frel}) the relative phase distribution in
the form, corresponding to the case when $A$
 and $B$ vortices are situated at the points ${\bf R_A}=(x_A, 0)$ and
 ${\bf R_B}=(x_B, 0)$ and image $A$
 and $B$ antivortices are situated at the points ${\bf \tilde{R}_A}=(-x_A, 0)$ and
 ${\bf \tilde{R}_B}=(-x_B, 0)$ correspondingly:
 \begin{align}\label{grad-phi}
  &\nabla\varphi_{rel}=\\
  \nonumber
  &\frac{{\bf r}-{\bf R_{A}}}{|{\bf r}-{\bf R_{A}}|^2}-
  \frac{{\bf r}-{\bf \tilde{R}_{A}}}{|{\bf r}-{\bf
  \tilde{R}_{A}}|^2}-
  \frac{{\bf r}-{\bf R_{B}}}{|{\bf r}-{\bf R_{B}}|^2}+
  \frac{{\bf r}-{\bf \tilde{R}_{B}}}{|{\bf r}-{\bf
  \tilde{R}_{B}}|^2}.
 \end{align}
The result is
 \begin{align}\label{Frel2}
 & F_{nc}= \\
 \nonumber
 &\frac{\phi_{A} \phi_{B}}{(4\pi\lambda)^2}
 \left[\ln\left(\frac{2x_A}{\xi_A}\right)+\ln\left(\frac{2x_B}{\xi_B}\right)-2\ln\left|\frac{x_A+x_B}{x_A-x_B}\right|
 \right].
 \end{align}

 Note that the expression (\ref{Fm2}) for the energy $F_m$
 as well as the expression (\ref{Frel2}) for $F_{nc}$ are valid
 only when the separations between vortices are larger than the
 coherence lengths: $x_A>\xi_A$, $x_B>\xi_B$ and
 $|x_A-x_B|> {\rm max} (\xi_A,\xi_B)$. However,
 the last restriction is removed when the total free energy
 $$
 F_{vv}(x_A,x_B)=F_{m}(x_A,x_B)+F_{nc}(x_A,x_B)
 $$
 is considered. Indeed, in this case the
 logarithmic singularities at $x_A=x_B$ in Eqs. (\ref{Fm2})
 and (\ref{Frel2}) cancel each other. Therefore the expression for the total free
 energy $F_{vv}(x_A,x_B)$ given by the sum of partial energies (\ref{Fm2})
 and (\ref{Frel2}) can be used for
 any values of the vortex coordinates $x_A> \xi_A$ and $x_B>
 \xi_B$.

Finally let us consider the second term in Eq.(\ref{Gibbs0}) which
determines the interaction of vortices with the external magnetic
field. This interaction energy can be written as
 \begin{equation}\label{InteractionEnergy}
 W=-\frac{H_0\Phi}{4\pi},
 \end{equation}
where $\Phi=\int H_v d^2{\bf r}$ is the total magnetic flux
generated by the vortex currents. Due to the linearity of
  the London-Maxwell equation (\ref{London-Maxwell1})
 and the boundary condition (\ref{BC}) the magnetic flux can be
 decomposed into the contributions from A and B vortices:
 $$
 \Phi=\Phi_A+\Phi_B.
 $$
 In accordance with the consideration of the
 Ref.(\onlinecite{Schmidt}) these partial magnetic fluxes
  depend on the positions of fractional vortices in the following
  way:
 \begin{equation}\label{Flux-A}
 \Phi_A=\phi_{A} \left(1- e^{-x_A/\lambda}\right)
 \end{equation}
 \begin{equation}\label{Flux-B}
 \Phi_B=\phi_{B} \left(1- e^{-x_B/\lambda}\right).
 \end{equation}
  Thus for the energy of vortex interaction with the external
  magnetic field we finally obtain:
  $$
 W(x_A,x_B)=
 $$
 \begin{equation}\label{InteractionEnergyVV}
  -\frac{H_0}{4\pi} \left[\phi_{A}
  \left(1- e^{-x_A/\lambda}\right)+\phi_{B}\left(1-
  e^{-x_B/\lambda}\right)\right].
 \end{equation}

 \subsection{Stable fractional vortices}
 \label{subsec:StableVortices}

Summing up the different parts of the free energy
(\ref{Fm2},\ref{Frel2}) and the energy of interaction with the
external magnetic field (\ref{InteractionEnergyVV}) we obtain the
Gibbs energy
 \begin{equation}\label{Gibbs-molecule}
 F_G(x_A,x_B)=F_{vv}(x_A,x_B)+W(x_A,x_B).
 \end{equation}
 The equilibrium vortex
  positions are determined by the extremum of the Gibbs energy,
  which is given by the condition
 \begin{align}\label{Extremum1}
 & \frac{\partial F_G}{\partial x_A}(x^*_A,x^*_B)=0\\
  \nonumber
 & \frac{\partial F_G}{\partial x_B}(x^*_A,x^*_B)=0
 \end{align}
 if both $x^*_A>\xi_A$, $x^*_B>\xi_B$ or otherwise by the
 condition
  \begin{align}\label{Extremum2}
 & \frac{\partial F_G}{\partial x_A}(x^*_A,\xi_B)=0\\
 \nonumber
 &\frac{\partial F_G}{\partial x_B}(\xi_A,x^*_B)=0
 \end{align}
 when only one fractional vortex enters the superconductor.

\begin{figure}[htb]
\centerline{\includegraphics[width=1.0\linewidth]{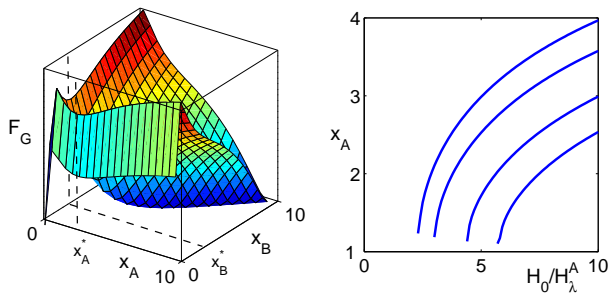}}
\caption{\label{Fig:Energy+Extremum} {\it Left panel}: Gibbs
energy of the molecule consisting of two fractional vortices near
the surface of superconductor. The parameters are $\xi_A=\xi_B$
and $\phi_A=\phi_B$. The magnetic field is taken $H_0=5H_\lambda$.
In this plot the local energy minima are clearly seen at $x_B=0$,
$x_A=x^*_A$ and $x_A=0$, $x_B=x^*_B$. {\it Right panel}: The
dependence of the fractional vortex position $x^*_A$ on the
magnetic field $H_0$ for different values of $\phi_A/\phi_B =
1,2,3,4$ (from top to bottom curve). All lengths are normalized to
the scale $\lambda$.
  }
\end{figure}

 As it is shown in left panel of Fig.(\ref{Fig:Energy+Extremum}) for
 sufficiently large magnetic field there can be realized a situation when
 the fractional vortices are stable near the surface. In general it corresponds to the extremum condition
 (\ref{Extremum2}) which means that fractional vortices of different types can not coexist, i.e. either $x^*_B=0$ and
 $x^*_A\neq 0$ or $x^*_A=0$ and $x^*_B\neq 0$. Another alternative is given by the condition (\ref{Extremum1})
 which is satisfied only for a composite
  vortex with $x^*_B=x^*_A\neq 0$, but it is impossible to realize a regime of stable molecule
 of fractional vortices near the surface with $x^*_B\neq x^*_A\neq 0$.

  Let us now discuss in more detail the dependence of the
 equilibrium position of a fractional vortex on the
 external magnetic field ${\bf H_0}=H_0 {\bf z_0}$. To be definite we consider the
 case of $A$ phase vortex.  Then the equation which determines
 $x^*_A$ follows from the condition (\ref{Extremum2}) and has the
 following form
 \begin{equation}\label{FgDerivative}
  \left[ K_1 (y) + \frac{1}{\sigma y}\right] e^{y/2} =
  2\pi\frac{H_0}{H^A_{\lambda}},
 \end{equation}
 where  $y=2x^*_A/\lambda$ and $H^A_\lambda=\phi_A/(2\pi\lambda^2)$.
 The solution of this equation is plotted in right panel Fig. (\ref{Fig:Energy+Extremum}) as a function of external
 magnetic field $H_0$ for the different values of coefficient $\sigma=1,2,3,4$ (from top to
bottom curve). Several qualitative features which characterize the
behaviour of fractional vortices near the surface can be deduced
from the Fig. (\ref{Fig:Energy+Extremum})b. Firstly, the distance
$x^*_A$ is scaled in the penetration depth $\lambda$. This length
scale is determined by the decay length of the interaction of
vortex with the external field, i.e. the energy $W(x_A)$.
Secondly, the vortex coordinate $x_A^*$ grows monotonically with
the external field $H_0$. Such behaviour is explained by the fact
that the force pushing the vortex towards the bulk of
superconductor is proportional to $H_0$. Thus the larger is the
pushing force the farther fractional vortex can penetrate into
superconductor. Finally, the minimal magnetic field which provide
the local Gibbs energy minimum for $x^*_A\neq 0$ is of the order
of $H^A_\lambda=\phi_A/(2\pi\lambda^2)$.

 As we have shown above if the external magnetic field
 is larger than the threshold value of the order of
 $H^i_\lambda=\phi_i/(2\pi\lambda^2)$, where $i=A,B$ there appears
 a minimum of the fractional vortex energy which determines the
 equilibrium vortex position $x^*_i> 0$. The Gibbs energy
 minima occur both for the $A$ and $B$ vortices
 for an arbitrary ratio of the fractional fluxes
 $\phi_A$ and $\phi_B$ as well as of the coherence lengths $\xi_A$ and $\xi_B$.
 For example in the left panel of Fig.(\ref{Fig:Energy+Extremum}) the energy minima
  are shown to exist at $x_A^*\neq 0$, $x_B^*=0$ and $x_B^*\neq 0$, $x_A^*=0$
  for a two gap superconductor with $\phi_A=\phi_B$
  and $\xi_A=\xi_B$.

 Thus we can conclude that fractional vortices can be stable near the surface
 of superconductor for  arbitrary values of parameters $\phi_A, \xi_A$
 and $\phi_B, \xi_B$ characterizing the two
 superconducting condensates. Still the consideration above is not
 complete because it misses a very important question of how the
 fractional vortices can be created in the superconducting region.
   In the next section we will see that the scenario of vortex penetration is
  sensitive to the ratio of the coherence lengths $\xi_A$ and $\xi_B$
  of two different condensates.

 \subsection{Bean-Livingston barrier and the critical field of the first vortex entry}
 \label{subsec:CriticalField}

  To begin with the analysis of vortex penetration we
  note at first that the condition for a Bean-Livingston barrier
  suppression derived in Ref.\onlinecite{Schmidt} should be modified here to
 take into account the peculiarities of a two gap superconductor case.
 This condition takes the following form:
 \begin{equation}\label{Eq:PenetrationCondition}
  ({\bf n}\cdot\nabla F_G)(x_A=\xi_A,x_B=\xi_B)<0
 \end{equation}
 for some vector ${\bf n}=(n_A,n_B)$ satisfying the condition
 $n_A>0$, $n_B>0$. With the help of the expression for the Gibbs energy
 given by the Eqs.(\ref{Fm2}, \ref{Frel2}, \ref{InteractionEnergyVV}, \ref{Gibbs-molecule})
 it is possible to obtain an analytical expression for the critical value of magnetic field $H_0=H_s$
 suppressing the surface barrier.

It occurs that the two different regimes of vortex entry are
determined by the ratio of coherence lengths of two
superconducting components $\xi_A$ and $\xi_B$.

\begin{figure}[htb]
\centerline{\includegraphics[width=1.0\linewidth]{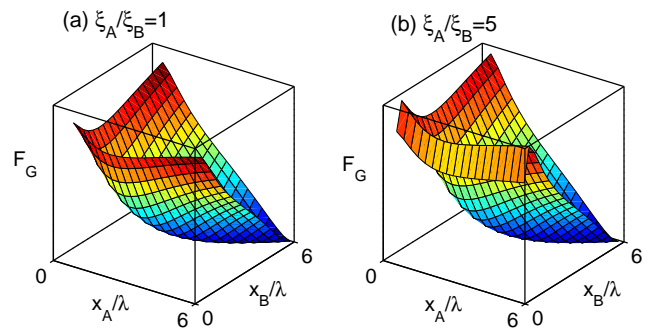}}
\caption{\label{2Denergy} Gibbs energy of the molecule consisting
of two fractional vortices near the surface of superconductor. (a)
Equal coherence lengths of two condensates  $\xi_A/\xi_B=1$. In
this case the Bean-Livingston barrier is suppressed simultaneously
for fractional vortices of both types. (b) Different coherence
lengths of two condensates $\xi_A/\xi_B=5$. In this case the
barrier is suppressed for the fractional vortex in $A$ condensate
while for the $B$ condensate it still exists. }
\end{figure}

{\bf (i)} The case when $\xi_A=\xi_B=\xi$ always corresponds to
the simultaneous entrance of two fractional vortices. As one can
see in Fig.(\ref{2Denergy})a the surface barriers for two vortices
disappear at the same value of the critical field $H_0=H_s$ which
is estimated as
 \begin{equation}\label{Eq:Hs0}
 H_s=\frac{\phi_0}{4\pi\lambda\xi}.
 \end{equation}

{\bf (ii)} The case when $\xi_A\neq \xi_B$ is more interesting.
 Indeed in Fig.(\ref{2Denergy})b it is clearly shown that there is
 a range of magnetic fields when for one of the superconducting condensates
 the Bean-Livingston barrier is already suppressed while for the other it still exists.
 The critical field of the first vortex entry $H_0=H_{s1}$ is given by
 \begin{equation}\label{Eq:Hs1}
 H_{s1}={\rm min} \left(H_{sA}, H_{sB}\right),
 \end{equation}
  where $H_{sA}=\phi_0/(4\pi\lambda\xi_A)$ and
  $H_{sB}=\phi_0/(4\pi\lambda\xi_B)$.
 Therefore the vortex with larger core size $\xi_A>\xi_B$ is the first one to penetrate the
 superconductor.

  Thus we obtain that in some range of
 parameters it is possible that vortices penetrate by parts, i.e.
  only the fractional vortices in one of the condensates appear at first in the superconductor.
  Such situation is possible when the coherence lengths of
  the two condensates are different $\xi_A\neq \xi_B$. In this case the vortex penetration
  occurs according to the following scenario. If the external
  magnetic $H_0$ field is increased from zero then at the
  threshold value $H_0=H_{s1}$ the Bean-Livingston barrier is
  suppressed and the fractional vortices with larger core size
  enter the superconductor. However, they can not proliferate into
  the bulk and sit at some equilibrium position near the surface
  of superconductor. The equilibrium position of one fractional vortex
  is determined by the Eq.(\ref{FgDerivative}) and its dependence
  on the external field is shown in the right panel of Fig.\ref{Fig:Energy+Extremum}.
  When the magnetic field $H_0$ is increased further and
  reaches the second threshold field $H_0=H_{s2}$ determined by
 \begin{equation}\label{Eq:Hs2}
 H_{s2}={\rm max} \left(H_{sA}, H_{sB}\right),
 \end{equation}
  the surface barrier is suppressed for the vortices in the second
  condensate. As one can see from Fig. (\ref{2Denergy}) the
  fractional vortices in $A$ and $B$ condensates always merge
  to form a composite vortex and proliferate into the bulk superconductor
    since the total minimum of the Gibbs energy is always reached at
    $x^*_A=x^*_B$. Therefore we can conclude that if $\xi_A\neq\xi_B$ then at the
    range of magnetic fields $H_{s1}<H_0<H_{s2}$
    the fractional vortices should appear near the surface of a
    two-gap superconductor.

    Note that above we have analyzed only the single
    vortex problem. However, it is natural to expect that when the
    Bean-Livingston barrier for fractional vortices is suppressed
    they penetrate into the superconductor until some equilibrium
    vortex distribution is set up near the surface. Below we will find the
    equilibrium distribution of fractional vortices at the range
    of magnetic fields $H_{s1}<H_0<H_{s2}$.

\subsection{Unconventional magnetic state with finite
density of fractional vortices near the surface.}
\label{subsec:chains}

Now we proceed with considering the unconventional magnetic state
of two-component superconductor which is realized when the finite
density of stable fractional flux vortices appears near the
surface as the magnetic field exceeds the critical field of the
first vortex entry $H_{s1}$. To be definite we assume that
$\xi_A>\xi_B$ and therefore the vortex in
 $A$ phase is the first one to enter the superconductor since
$H_{sA}<H_{sB}$.

In general the task of finding the equilibrium configuration of
fractional vortices stabilized near the surface of superconductor
seems to be a rather complicated one. One of the reasons is a
many-body origin of this problem, resulting in a huge number of
 dimensions of a configuration space. However we can use here a
 simplified approach which is based on the consideration of a
 distribution of an average density of fractional vortices. Such
 treatment is justified by the fact that in a magnetic field of the
 order $H_{sA}$ the intervortex distance is of the order of $a\sim \sqrt{\lambda\xi_A}$.
  As we have shown above the fractional vortices sit
 at the distance $d\sim\lambda$ from surface, which is much larger
 than the characteristic intervortex distance.

 So we assume that the $A$ phase vortices are distributed in space
 with the density $n({\bf r})$, where ${\bf r}=(x,y)$. To find the form of the equilibrium
 vortex distribution $n({\bf r})$ we should consider a probe
 vortex positioned at some point ${\bf r_v}=(x_v,y_v)$ and place a
 requirement of its stationarity, i.e. that the total force acting
 on the probe vortex is to be zero
 $$
  f=\frac{\partial}{\partial x_v} F_G (x_v)=0.
 $$

 The expression for
 the Gibbs energy consists of three parts, namely
 $$
 F_G(x_v)=F_m (x_v)+F_{nc} (x_v)+W (x_v),
 $$
 given by the Eqs. (\ref{Fm}, \ref{Frel}, \ref{InteractionEnergy})
 correspondingly. So to calculate the force we should determine the dependence of
 three energy parts on the position of the probe vortex.
 This leads to the following integral equation for the vortex
 density $n(x)$ (the details of calculation are given in the
 Appendix \ref{Appendix2}):
 \begin{align}\label{IntegralEquation0}
  & \alpha\int_{0}^{\infty}\left[ \frac{(x-x_v)}{|x-x_v|} e^{-|x-x_v|}
  + e^{-|x+x_v|}\right]n(x) dx+\\
  \nonumber
  & \beta  \int_{x_v}^{\infty} n(x)dx = \gamma  e^{-x_v},
 \end{align}
   where
   $\alpha=\phi^2_A/(16\pi\lambda)$,
   $\beta=\phi_A\phi_B/(8\pi\lambda)$ and
   $\gamma=\phi_A H_0/(8\pi\lambda)$. In Eq.(\ref{IntegralEquation0})
   the coordinates $x$ and $x_v$ are normalized to the length scale $\lambda$.

  The integral Eq.(\ref{IntegralEquation0}) can be solved
  analytically.
 Differentiating three times the both sides of Eq.(\ref{IntegralEquation0})
  over $x_v$ we obtain the differential equation for $n(x)$
 \begin{equation}\label{DifferentialEquation1}
  \frac{d^2 n}{dx^2} = \frac{\beta}{\beta+2\alpha} n.
 \end{equation}
   Therefore the solution $n(x)$ has the form
 \begin{equation}\label{Ansatz}
 n(x)=A e^{- x/L_n},
 \end{equation}
 where the scale $L_n$ and the amplitude $A$ are determined by the
  Eqs.(\ref{IntegralEquation0},\ref{DifferentialEquation1}) as
  follows (here we restore the length normalization factor $\lambda$):
 \begin{equation}\label{ParameterKappa}
 L_n=\lambda\sqrt{\frac{2\alpha+\beta}\beta{}}=
 \lambda\sqrt{1+\phi_A/\phi_B}
 \end{equation}
 \begin{equation}\label{ParameterA}
  A=\frac{\gamma}{\alpha}\frac{(L^2_n-\lambda^2)}{2 L^2_n}=\frac{H_0}{\phi_A}\frac{(L^2_n-\lambda^2)}{L^2_n}.
 \end{equation}

Thus we obtain that the concentration of fractional vortices
$n(x)$ given by
Eqs.(\ref{Ansatz},\ref{ParameterKappa},\ref{ParameterA}) decays
exponentially with the distance from the surface of
superconductor. Note that in case of ordinary vortices (which is
obtained by putting $\beta=0$ in the expressions above)\ we get
that $L_n=\infty$, which means that the vortex density is
constant\cite{Schmidt}. The difference in the behaviour of
ordinary and fractional vortex density $n(x)$ is provided by an
additional force which is determined by the gradient of energy
$F_{nc}(x)$ (\ref{Frel}). This force pushes the fractional
vortices out of superconductor. Qualitatively this force is
explained by the fact that the existence of fractional vortices is
not allowed in the bulk superconductor due to the infinite energy
of such objects.

Let us find a contribution of fractional vortices to the average
 magnetization of a superconducting sample. To distinguish from the magnetization
 provided by Meissner currents we denote this contribution as
${\bf M_f}=M_f {\bf z_0}$. Then we have:
 \begin{equation}\label{MagneticMoment0}
 M_f=\frac{1}{4\pi L}\int_0^\infty n(x)\Phi_A(x) dx,
 \end{equation}
 where $n(x)$ is the vortex density, $L$ is the sample size in $x$ direction
  (see Fig.\ref{Fig:model}) and $\Phi_A (x)$ is the total magnetic
 flux (\ref{Flux-A}) provided by one fractional vortex
placed at the distance $x$ from the boundary.
Substituting expression (\ref{Ansatz}) for the vortex density into
Eq.(\ref{MagneticMoment0}) we obtain that
 \begin{equation}\label{MagneticMoment1}
 M_f=\frac{H_0}{4\pi} \frac{L_n-\lambda}{L}.
 \end{equation}

\section{Discussion}
\label{sec:discussion}

 Now with the help of the results of previous section
 let us consider the properties of the magnetization
 curve of a two-gap superconductor. The qualitative features
 of the magnetization should be determined by the following
  facts.

 {\bf (i)} In increasing magnetic field vortices penetrate into superconductor
 by parts. When the magnetic field reaches the critical
 value (\ref{Eq:Hs1}) the fractional vortices
 with larger core size proliferate into superconducting region.

{\bf (ii)} For the range of magnetic fields $H_{s1}<H_0<H_{s2}$,
 only fractional vortices of one type exist near the surface of superconductor.
Their distribution is determined by
Eqs.(\ref{Ansatz},\ref{ParameterKappa},\ref{ParameterA}). When the
magnetic field becomes larger $H_0>H_{s2}$ the fractional vortices
of the other type penetrate. Then fractional vortices of different
types merge and composite vortices proliferate into the bulk
superconductor. The value of the field $H_{s2}$ is certainly
modified by the presence of finite density of fractional vortices
in the first condensate. The modified value of $H_{s2}$ can be
found by calculating the forces  forces acting of a single
fractional vortex in the second condensate in a same way as on
that in the first condensate (see Appendix\ref{Appendix2}). Then
we obtain that in this case the critical field is reduced by the
factor $\sqrt{\phi_B/\phi_0}$ as compared to Eq.(\ref{Eq:Hs2}),
where $\phi_B$ is a magnetic flux carried by fractional vortices
in the second condensate.

 Thus we can work out the following qualitative picture of
 the superconducting sample magnetization behaviour in increasing magnetic field.
 At low fields $H_0<H_{s1}$ the magnetization is determined by the Meissner
 current. If all the stray fields are neglected it is given by
 \begin{equation}\label{M1}
  M_1=-\frac{H_0}{4\pi}.
 \end{equation}

 For magnetic fields from the range $H_{s1}<H_0<H_{s2}$ the magnetization is changed due to the
 fractional vortices
 \begin{equation}\label{M2}
  M_2=M_f(H_0)-\frac{H_0}{4\pi},
 \end{equation}
  where $M_f(H_0)$ is given by Eq.(\ref{MagneticMoment1}).
 Finally at $H_0>H_{s2}$ the magnetization changes abruptly and becomes
 \begin{equation}\label{M3}
  M_3=\frac{B(H_0)-H_0}{4\pi},
 \end{equation}
  where the net magnetic induction $B(H_0)$ is
determined by the configuration of the vortex lattice in bulk
superconductor (see for example
Refs.\onlinecite{Schmidt},\onlinecite{deGennes}).

\begin{figure}[htb]
\centerline{\includegraphics[width=1.0\linewidth]{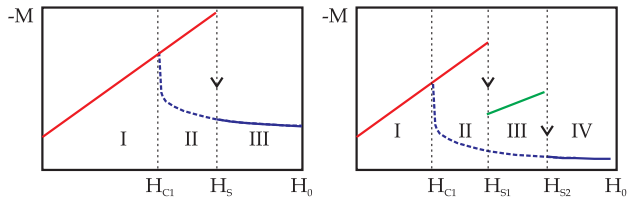}}
\caption{\label{Fig:Magnetization} Qualitative behaviour of
magnetization curve of a superconductor modified due to the
surface barrier. {\it Left panel}: Single gap superconductor. {\it
Right panel}: Two-gap superconductor. }
\end{figure}

These qualitative features of the magnetization curve are
summarized in Fig.(\ref{Fig:Magnetization}), where the magnetic
field dependence of the average magnetization of superconducting
sample $M(H_0)$ is shown for the case of single gap (left panel)
and two gap (right panel) superconductors. For a single-gap
superconductor we have a conventional picture of the Meissner
state overheating due to the Bean-Livingston barrier. In the left
panel of Fig.(\ref{Fig:Magnetization}) we show the three
characteristic regions corresponding to this case. At the region
(I) when  the magnetic field is smaller than the first critical
one $H_{c1}$ the Meissner state is realized. At the region (II)
where $H_{c1}<H_0<H_s$, where $H_s=\phi_0/(4\pi\lambda\xi)$ the
Meissner state is overheated due to the Bean-Livingston barrier.
Correspondingly, the total magnetization still grows at this
region (solid line) instead of following the bulk sample curve
(dash line). Finally at $H_0=H_s$ the surface barrier is
suppressed and the magnetization jumps to the value determined by
the configuration of vortices in the bulk of superconductor.

 In contrast to the conventional scheme a two gap superconductor features two jumps
 of the magnetization curve [right panel of Fig.(\ref{Fig:Magnetization})].
 Correspondingly there are four regions which are characterized by
 qualitatively different behaviour of the $M(H_0)$ dependence. The
 regions (I) where $H_0<H_{c1}$ and (II) where
 $H_{c1}<H_0<H_{s1}$ are the same as for
 conventional superconductor. A nontrivial behaviour starts
 at the field $H_0=H_{s1}$ of the first vortex entry. At this threshold
 field value there is a jump of the magnetization curve,
 which is determined by a set up of a finite concentration of
 fractional vortices near the surface of superconductor.
 At the region (III) for the range of magnetic fields $H_{s1}<H<H_{s2}$ the
 magnetization  is determined by the Eqs. (\ref{MagneticMoment1},\ref{M2}) and grows linearly
 with $H_0$. The second jump of the $M(H_0)$ curve occurs at the field $H_0=H_{s2}$
 when the composite vortices start to proliferate into the bulk of
 superconductor. Finally at the region (IV) when $H_0>H_{s2}$ the
 value of magnetization is  determined by the configuration of
 composite vortices in the bulk of superconductor.

 The discussed above possibility of separate vortex penetration in
 two coexisting superconducting condensates should be especially
 interesting in connection with the investigation of projected
 superconducting state of liquid metallic hydrogen
 \cite{Hydrogen1,Hydrogen2}, where the superconducting state is
 formed by electronic and protonic Cooper pairs. The observation
 of the two-component superconducting state in this case cannot be
 implemented by conventional techniques and requires special
 experimental approaches\cite{Hydrogen3}. In particular, the most
 challenging problem of the protonic superconductivity detection
 cannot be treated by the standard measurement of the Meissner
 effect since the critical temperature for electrons is estimated
 to be much larger than that of protons. Therefore, in the Meissner state
 the contribution to the total magnetic moment of protonic
 supercurrent will be always masked by that of the electronic
 component. On the other hand, as we have shown above the
 relaxation of the overheated Meissner state by vortex penetration
 should feature an additional jump of magnetization due to the
 coexistence of protonic and electronic superconducting
 components. According to our results the critical magnetic field
 of the first vortex entry is determined by the condensate with
 the largest coherence length, which is the one with the smallest critical
 temperature. In liquid metallic hydrogen such component is always a protonic one,
 therefore the first jump on the magnetization curve should be
 determined by the vortices in protonic superfluid provided it is a type-II one.
  According to the recent estimations the low temperature limit of
   magnetic flux carried by vortices in
 protonic component is of the order $\phi_p(T=0)\sim 10^{-3}\phi_0$
 being by several orders of magnitude larger than the
 resolution threshold in modern experiments\cite{Hydrogen3}
 which allow to detect the magnetic flux less then
 $10^{-5}\phi_0$. Therefore even for the temperatures close to the
 critical one of protinic component it is possible to detect the
 protonic vortices.
 It is interesting to note that on approaching the
 critical temperature of protonic component from below the
 penetration field for the fractional vortex in protonic condensate goes to zero.
 Therefore the largest interval of magnetic fields where only the fractional vortices
 should exist near the surface is in the vicinity of protonic critical
 temperature.

 Finally we note that besides an intriguing possibility to
 explore the nature of superconducting state in liquid metallic
 hydrogen our results are applicable to the case of conventional
 two-gap superconductor. Generally speaking in this case one should take into
 account the effect of interband Josephson coupling that we have
 neglected. This assumption is justified in case when the Josephson length
 of the intercomponent phase difference relaxation is much larger
 than the other relevant length scales, that are the coherence lengths
 in both condensates and the London penetration
 length. However even if this condition is violated,
 the critical magnetic field of the first vortex entry $H_{s1}$ remains the
 same, i.e. it is not affected by the presence of the Josephson
 coupling since the energy of Josephson string connecting the
 fractional vortex with the surface grows quadratically at small
 distances (see e.g. Ref.\onlinecite{JString}) and does not alter the condition of vortex penetration
 (\ref{Eq:PenetrationCondition}). The further penetration of vortex
 in the second component is certainly affected by the presence of
 the first fractional vortex emanating the Josephson string ending at the surface.
 Presumably, the result should be the reduction of the field $H_{s2}$ of the second vortex entry,
 however the strict quantitative investigation of this effect is
 beyond the scope of the present paper.
 Thus we can conclude that our main result, i.e. the separate penetration of fractional vortices and
 the presence of two jumps on the magnetization curve remains qualitatively relevant even in case of rather strong
 intercomponent Josephson interaction.

\section{Summary}
\label{sec:summary}
 To sum up we have investigated the unconventional magnetic state
 of two-component superconductors. This peculiar state is realized when
 the finite density of stable fractional vortices appears near the
 surface of superconductor under the action of external magnetic
 field. This result contrasts to the case of bulk two-component superconductor
 where fractional vortices have infinite energy and therefore can
 not exist\cite{Babaev-GL}. In our situation the stability of
 fractional vortices near the surface is provided by the
 cancellation of the unscreened superconducting current due to the
 image antivortices.

 Also we have discussed the influence
 of fractional vortices on a Bean-Livingston barrier and the
 magnetization curve of a two-component superconductor. In particular,
 we have found that if the coherence lengths of two condensates
 are different then vortices penetrate the superconductor
 by parts. The fractional vortices in the condensate with
 larger coherence length, i.e. having the larger core size are the
 first to enter the superconductor. By increasing the
 magnetic field further the fractional vortices in the second
 condensate are pushed to penetrate. Then the fractional vortices of two different types merge to
 form composite vortices which proliferate into the bulk superconductor.

  We have shown that the magnetization curve of a superconduting sample should feature
 the two jumps associated with the penetration of two types of
 fractional vortices. The observation of such a peculiar
 magnetization behaviour could be considered as an experimental
 identification of fractional vortices in systems with several superconducting
 components, such as multiband superconductors and the projected
 superconducting state of liquid metallic hydrogen.

\section{Acknowledgements}
It is my pleasure to thank Egor Babaev and Alexander Mel'nikov for
stimulating discussions. I am grateful for hospitality to the
Department of Theoretical Physics at The Royal Institute of
Technology where the final part of the work was done. This work
was supported by Swedish Research Council, ``Dynasty" Foundation,
Presidential RSS Council (Grant No.MK-4211.2011.2), Russian
Foundation for Basic Research, by Programs of RAS "Quantum Physics
of Condensed Matter" and "Strongly correlated electrons in
semiconductors, metals, superconductors and magnetic materials",
and by Russian Ministry of education under the Federal program
"Scientific and educational personnel of innovative Russia".

\appendix

 \section{Derivation of the integral equation (\ref{IntegralEquation0})}
\label{Appendix2}
 {\bf (i)} To find the energy $F_m(x_v)$ let us consider at first
 the interaction of the probe vortex with vortex positioned at the
 coordinate ${\bf r}=(x,y)$ and the antivortex at the coordinate ${\bf
 \tilde{r}}=(-x,y)$. Applying the Eq.(\ref{Fm1-0}) we get into account only the energy of interaction between vortices
 and obtain
 \begin{align}\label{B1}
 &F_m({\bf r_v},{\bf r})= \\
 \nonumber
 &2\left(\frac{\phi_A }{4\pi
 \lambda}\right)^2\left[K_0\left(\frac{|{\bf r}-{\bf
 r_v}|}{\lambda}\right)-
 K_0\left(\frac{|{\bf \tilde{r}}-{\bf
 r_v}|}{\lambda}\right)\right].
 \end{align}

 Now to find the total energy $F_m(x_v)$ we integrate over
 the positions of all vortices to obtain:
 \begin{equation}\label{B2}
  F_m(x_v)=\int_{-\infty}^{\infty} dy \int_{0}^{\infty} dx \; n(x) F_m({\bf r_v},{\bf r}).
 \end{equation}
 To perform the integration in Eq.(\ref{B2}) we use the following relation
 $$
 \int_{-\infty}^{\infty} K_0\left(\sqrt{x^2+y^2}\right) dy= \pi e^{-|x|}
 $$
 and obtain
\begin{align} \label{B3}
  F_m(x_v)= \frac{\phi^2_A }{8\lambda\pi} \int_{0}^{\infty}
   &\left[\exp \left(-\frac{|x-x_v|}{\lambda}\right)-\right. \\
 \nonumber
   &\left.\exp \left(-\frac{|x+x_v|}{\lambda}\right)\right]n(x) dx.
\end{align}

{\bf (ii)} According to the Eq. (\ref{Frel}) energy $F_{nc} (x_v)$
is determined by the following expression
 \begin{equation}\label{B4}
  F_{nc} (x_v)= \frac{1}{\pi}\frac{\phi_A\phi_B}{(4\pi\lambda)^2} \int d^2 {\bf r} \nabla \varphi \nabla
 \varphi_v,
 \end{equation}
 where
 $$
 \varphi_v=\arctan \left(\frac{y-y_v}{x-x_v}\right)
 $$
 is the phase distribution created by the probe vortex and
 $$
 \varphi_v=\sum_i m_i \arctan  \left(\frac{y-y_i}{x-x_i}\right)
 $$
 is the phase created by all other vortices ($m_i=1$) and
 antivortices ($m_i=-1$). To evaluate the expression
 (\ref{B4})
 let us consider again the interaction of the probe vortex with
 vortex positioned at the coordinate ${\bf r}=(x,y)$ and the antivortex at the coordinate ${\bf
 \tilde{r}}=(-x,y)$. Then we have that
 $$
  F_{nc} ({\bf r_v},{\bf r})=2\frac{\phi_A\phi_B}{(4\pi\lambda)^2} \ln\left(\frac{|{\bf r}+{\bf r_v}|}{|{\bf r}-{\bf
  r_v}|}\right).
 $$
 This expression is divergent at the point ${\bf r}={\bf r_v}$,
 which should be cut off at the vortex core size $\xi_A$. However, this singularity is
 integrable and we don't introduce this cutoff here.
 Then we should sum the contributions from all vortices according
 to the expression
 \begin{equation}\label{B5}
  F_{nc}(x_v)=\int_{-\infty}^{\infty} dy \int_{0}^{\infty} dx \; n(x) F_{nc}({\bf r_v},{\bf r}).
 \end{equation}
 The integration over the coordinate $y$ in Eq. (\ref{B5}) can be performed using
 the relation
 $$
 \int_{-\infty}^{\infty} \ln\left(\frac{|{\bf r}+{\bf r_v}|}{|{\bf r}-{\bf
  r_v}|}\right) dy=\left\{2\pi x_v \;\;\;{\rm for}\;\;\; x>x_v \atop
 2\pi x \;\;\;\;{\rm for}\;\;\; x<x_v \right.
 $$
 Then we obtain:
 \begin{align}\label{B6}
 &F_{nc}(x_v)=\\
 \nonumber
  &\frac{\phi_A\phi_B}{4\pi\lambda^2}
 \left[ \int_{0}^{x_v} x n(x)dx +
  \int_{x_v}^{\infty}x_v n(x)dx\right].
 \end{align}

 {\bf (iii)} Finally the energy $W (x_v)$ determining the
 interaction of the probe vortex with external magnetic field
  can be find from Eqs.(\ref{InteractionEnergy},\ref{Flux-A}):
 \begin{equation}\label{B7}
 W (x_v)= -\frac{\phi_A}{4\pi} H_0
 \left(1-e^{-x_v/\lambda}\right).
 \end{equation}

 Summing all the contributions to the Gibbs energy (\ref{B3},\ref{B6},\ref{B7}) we obtain the
 total force acting on a probe vortex:
  $$
 {\bf f}=\frac{\partial }{\partial {\bf r_v}} F_G=(f_m+f_{nc}+f_{ext}){\bf x_0},
 $$
 where
\begin{align}
 & f_m(x_v)= \frac{\phi^2_A }{8\pi\lambda^2} \int_{0}^{\infty}\left[ \frac{(x-x_v)}{|x-x_v|} \exp
 \left(-\frac{|x+x_v|}{\lambda}\right)\right. \\
 \nonumber
 & + \left.\exp \left(-\frac{|x+x_v|}{\lambda}\right)\right]n(x)
 dx,
\end{align}
 \begin{equation}
  f_{nc}=\frac{\phi_A\phi_B}{4\pi\lambda^2} \int_{x_v}^{\infty} n(x)dx
 \end{equation}
 and
 \begin{equation}
  f_{ext}=-\frac{\phi_A}{4\pi\lambda} H_0 e^{-x_v/\lambda}.
 \end{equation}
 The condition of the probe vortex stationarity ${\bf f}=0$ yields the
 integral Eq.(\ref{IntegralEquation0}), where the coordinates $x$
 and $x_v$ are normalized to the length scale $\lambda$.



\begin{thebibliography}{99}

\bibitem{MgB2}
T. Muranaka et al., {\it Frontiers in Superconducting Materials},
edited by A. V. Narlikar (Springer-Verlag, Berlin) (2005).

\bibitem{Ferropnictides}
V. Barzykin, L. P. Gor'kov, JETP Lett., {\bf 88}, 142 (2008); S.
Raghu, Xiao-Liang Qi, Chao-Xing Liu, D. J. Scalapino, and
Shou-Cheng Zhang, Phys. Rev. B {\bf 77}, 220503 (2008).

\bibitem{HeavyFermion}
M. Jourdan, A. Zakharov, M. Foerster, and H. Adrian, Phys. Rev.
Lett. {\bf 93}, 097001, (2004); G. Seyfarth, J. P. Brison, M.-A.
Measson, J. Flouquet, K. Izawa, Y. Matsuda, H. Sugawara, and H.
Sato, Phys. Rev. Lett. {\bf 95}, 107004 (2005).

\bibitem{Borocarbides}
S. V. Shulga, S.-L. Drechsler, G. Fuchs, K.-H. Muller, K. Winzer,
M. Heinecke, and K. Krug, Phys. Rev. Lett. {\bf 80}, 1730 (1998).

\bibitem{Experiment}
M. I. Eremets, I. A. Trojan, S. A. Medvedev, J.S. Tse, Y. Yao,
Science, {\bf 319}, 1506 (2008); S. Deemyad and I. F. Silvera,
Phys. Rev. Lett. {\bf 100}, 155701 (2008).

\bibitem{Hydrogen1}
N.W. Ashcroft, J. Phys. Condens. Matter {\bf 12}, A129 (2000);
Phys. Rev. Lett. {\bf 92}, 187002 (2004); K. Moulopoulos and N. W.
Ashcroft, Phys. Rev. Lett. {\bf 66} 2915 (1991).

\bibitem{Hydrogen2}
E. Babaev, A. Sudb{\o} and N.W. Ashcroft, Nature {\bf 431} 666
(2004).

\bibitem{Hydrogen3}
E. Babaev, A. Sudb{\o} and N.W. Ashcroft, Phys. Rev. Lett. {\bf
95} 105301 (2005).

\bibitem{Moskalenko} V.A. Moskalenko, Phys. Met. Metallogr. {\bf 8}, 503
(1959).

\bibitem{Suhl}
H.Suhl, B.T.Matthias, and L.R. Walker, Phys. Rev. Lett. {\bf 3},
552 (1959).

\bibitem{Babaev-VorticesNature}
E. Babaev and N.W. Ashcroft, Nature Physics {\bf 3}, 530 (2007).

\bibitem{Babaev-SemiMeissner}
E. Babaev, M. Speight, Phys. Rev. B {\bf 72}, 180502 (2005).

\bibitem{Moshchalkov1-5}
V. Moshchalkov, M. Menghini, T. Nishio, Q.H. Chen,A.V. Silhanek,
V.H. Dao, L.F. Chibotaru, N.D. Zhigadlo, and J. Karpinski, Phys.
Rev. Lett. {\bf 102}, 117001 (2009).

\bibitem{Babaev-GL}
E.Babaev, Phys. Rev. Lett. {\bf 89}, 067001 (2002).

\bibitem{VolovikRMP}
M.M. Salomaa and G.E. Volovik, Rev. Mod. Phys. {\bf 59}, 533
(1987).

\bibitem{ColdAtoms} R. Barnett, A. Turner, and E. Demler,
    Phys. Rev. A {\bf 76}, 013605 (2007).

\bibitem{HTSC} A.J. Berlinsky, A. L. Fetter, M. Franz, C. Kallin, and P. I.
Soininen, Phys. Rev. Lett. {\bf 75}, 2200 (1995); M. Franz, C.
Kallin, P. I. Soininen, A. J. Berlinsky, and A. L. Fetter, Phys.
Rev. B {\bf 53}, 5795 (1996); A. S.Mel'nikov, I. M. Nefedov, D. A.
Ryzhov, I. A. Shereshevskii, and P. P. Vysheslavtsev, Phys. Rev. B
{\bf 62}, 11 820 (2000).

\bibitem{ThFluct1}
E. Smorgrav, J. Smiseth, E. Babaev, and A. Sudb{\o},
 Phys. Rev. Lett. {\bf 94}, 096401 (2005).

\bibitem{ThFluct2}
J. Goryo, S. Soma and H. Matsukawa, Eur. Phys. Lett. {\bf 80},
 17002 (2007).

\bibitem{Babaev-NuclPhys}
E. Babaev, Nucl. Phys. B {\bf 686}, 397–412 (2004).

 \bibitem{Chibotaru}
 L. F. Chibotaru, V. H. Dao and A. Ceulemans,
 Eur. Phys. Lett. {\bf 78}, 47001 (2007).

 \bibitem{Peeters}
 R. Geurts, M. V. Milosevic, and F. M. Peeters, Phys. Rev. B {\bf
 81}, 214514 (2010).

\bibitem{NeutronStar1}
 P.B. Jones, Mon. Not. R. Astron. Soc. {\bf 371}, 1327–1333
 (2006).

\bibitem{NeutronStar2}
E. Babaev, Phys. Rev. Lett. {\bf 103}, 231101 (2009).

\bibitem{BeanLivingston}
C.P. Bean and J.D. Livingston, Phys. Rev. Lett. {\bf 12}, 14
(1964).

\bibitem{Schmidt}
 V.V. Schmidt, G.S. Mkrtchyan, Sov. Phys. Uspekhi {\bf 17}, 170
(1974).

\bibitem{Schopohl}
C. Iniotakis, T. Dahm, and N. Schopohl, Phys. Rev. Lett. {\bf
100}, 037002 (2008); A. Zare, T. Dahm, and N. Schopohl, Phys. Rev.
Lett. {\bf 104}, 237001 (2010).

\bibitem{Babaev-delocalization}
E. Babaev, J. J\"aykk\"a, and M. Speight, Phys. Rev. Lett. {\bf
103}, 237002 (2009).

\bibitem{deGennes}
 P.G. de Gennes, {\it Superconductivity of Metals and Alloys},
Benjamin, New York, (1966).

\bibitem{Abramoviz}
M. Abramowitz and I.A. Stegun, eds. (1972), {\it Handbook of
Mathematical Functions with Formulas, Graphs, and Mathematical
Tables}, New York: Dover Publications.

\bibitem{JString}
Y. Y. Goldschmidt and S. Tyagi,  Phys. Rev. B {\bf 71}, 014503
(2005).

\end{thebibliography}
\end{document}